\documentclass{article}
\newcommand{\be}{\begin{equation}} 
\newcommand{\ee}{\end{equation}} 
 
\def\simlt{\stackrel{<}{{}_\sim}} 
\def\simgt{\stackrel{>}{{}_\sim}} 
\def\bea{\begin{eqnarray}} 
\def\eea{\end{eqnarray}} 
\def\bean{\begin{eqnarray*}} 
\def\eean{\end{eqnarray*}} 
 
\newcommand{\barr}{\begin{array}} 
\newcommand{\earr}{\end{array}} 
 
\newcommand{\bed}{\begin{displaymath}} 
\newcommand{\eed}{\end{displaymath}} 
\newcommand{\bal}{\begin{array}{ll}} 
\newcommand{\eal}{\end{array}} 
 
\def\bvec#1{\raise1.5ex\hbox{$\rightarrow$}\mkern-16.5mu #1}

\begin{document}

\pagestyle{empty}

\rightline{UFIFT-HEP-05-27}
\rightline{hep-ph/0510256} \vspace*{1.5cm}

\begin{center}

\LARGE{Viewing Lepton Mixing through the Cabibbo Haze\\[20mm]}

\large{Lisa L. Everett\\[8mm]}

\it{Institute for Fundamental Theory, Department of Physics\\ University 
of Florida, Gainesville, FL, 32611, USA\\[5mm]}
 
\large{\rm{Abstract}} \\[7mm]

\end{center}

\begin{center}

\begin{minipage}[h]{14cm} 
We explore the hypothesis that the Cabibbo angle is an expansion parameter 
for lepton as well as quark mixing.  Cabibbo effects are deviations from 
zero mixing for the quarks but are deviations from unknown mixings for the 
leptons, such that lepton mixing is veiled by a Cabibbo haze.  
We present a systematic classification of parametrizations and catalog
the leading order Cabibbo effects.  We find that the size of the CHOOZ 
angle is not always correlated with the observability of CP violation. This 
phenomenological approach has practical merit both as a method for 
organizing top-down flavor models and as a guideline for planning future 
experiments.

\end{minipage}

\end{center}

\newpage

\pagestyle{plain}

\section{Introduction}
The observation of neutrino oscillations provides concrete evidence 
that neutrinos are massive and lepton mixings are observable.  Current 
observations do not allow for an unambiguous extraction of the neutrino 
mass pattern. However, the Maki-Nakagawa-Sakata-Pontecorvo (MNSP) lepton 
mixing matrix \cite{Maki:1962mu,Pontecorvo:1957cp} has now been measured 
\cite{salt,Aharmim:2005gt,sksolar,skatm,kamland,chooz,k2k}, revealing 
fundamental differences between the lepton and quark mixings. The quark 
mixing angles of the Cabibbo-Kobayashi-Maskawa (CKM) matrix are all small, 
with the largest angle given by the Cabibbo angle $\theta_c\simeq 
13^\circ$.  In contrast, two of the lepton mixing angles are large (the 
atmospheric and solar angles) and one is small (the CHOOZ angle, which is 
not yet measured but bounded to be $\simlt \theta_c$ \cite{chooz}). The 
challenge to understand this discrepancy provides an intriguing framework 
in which to explore the flavor puzzles of the SM, particularly within the 
context of quark-lepton grand unification \cite{patisalam}, for which all 
available data can be synthesized in the search for a credible theory of 
flavor.

As a step toward this goal, recently a phenomenological approach 
was advocated in which parametrizations of the lepton mixing matrix were 
developed as an expansion in $\lambda\equiv\sin\theta_c\simeq0.22$ 
\cite{Datta:2005ci} in analogy with Wolfenstein's 
parametrization of the CKM \cite{wolfenstein}. 
In addition to its practical advantages for phenomenology, 
the Wolfenstein parametrization hints at a guiding principle for flavor 
theory by providing a framework for examining quark mixing in the 
$\lambda \rightarrow 0$ limit.  Extending this hypothesis to the lepton 
sector, one finds that the lepton mixings are {\it unknown} in the 
$\lambda \rightarrow 0$ limit (unlike the quark mixings, which vanish).  
Hence, if the limit of zero Cabibbo angle is
meaningful for theory, there is a {\it Cabibbo haze} \cite{Datta:2005ci} 
in lepton mixing, in which the initial or ``bare" values of the 
mixings are screened by Cabibbo-sized effects.

In this approach, Cabibbo effects can be deviations from zero mixing (as 
in the quark sector) or deviations of the 
mixing angles from presumably large initial values.  Parametrizations of 
the MNSP matrix are categorized according to the bare mixings and the 
structure of the allowed perturbations.  
Perturbations which are linear in $\lambda$ yield shifts of 
$\simlt \theta_c\simeq 13^\circ$ (or larger if multiple 
contributions shift the same parameter), while ${\cal O}(\lambda^2)$ 
shifts are $\sim 3^\circ$.  CP violating phases can enter the ${\cal 
O}(\lambda)$ shifts but may only occur at subleading order, in which case 
the effective MNSP phase is suppressed and the size of 
the CHOOZ angle does not dictate the size of CP-violating observables.
Cabibbo haze was previously explored 
for the theoretically motivated class of models with a vanishing initial 
CHOOZ angle \cite{Datta:2005ci}.  The purpose of this paper is to 
generalize this 
approach to additional models and to provide a complete catalog of 
the dominant ${\cal O}(\lambda)$ effects.

Cabibbo haze allows for many Wolfenstein-like parametrizations of 
the MNSP matrix which are consistent with current data, although 
regularities should emerge upon more precise measurements.
Despite these ambiguities, this 
classification scheme is worthwhile even at this stage of experimental 
observations.  Ultimately, it serves as an organizing principle for 
categorizing the many top-down flavor models based on a $\lambda$ 
expansion.  Using the Cabibbo angle as an expansion parameter for 
lepton mixing also provides a guideline for planning future experiments.  
Of particular interest is the program to 
measure the CHOOZ angle, for which future facilities are 
expected to reach the ${\cal O}(\lambda^2)$ range 
\cite{nufact}.  Within this general 
theoretical framework, probing the CHOOZ angle at this level will 
provide important insight into the nature of lepton mixing in 
the $\lambda\rightarrow 0$ limit.

This paper is structured as follows.  In Section 2, after a brief 
discussion of the current status of the data, we review the lessons from 
grand unification which form the basis of this approach.  We discuss the 
systematics of our theoretical framework in Section 3, and turn to a 
discussion of the leptonic Cabibbo shifts in the mixing angles and the 
CP-violating phases in Section 4. Finally, we present the conclusions and 
outlook.

\section{Preliminaries} 
We provide here a brief review of the data as well as the 
basic philosophy and motivation for our phenomenological approach.
We begin with the lepton data.\footnote{In 
this paper, we do not include constraints from the LSND 
reactor experiment \cite{Athanassopoulos:1997pv}, which necessitate either 
additional neutrino families and/or CPT violation.} 
In the standard parametrization (\cite{pdg} and references therein):
\bea 
\label{umnspstandardparam}
\mathcal{U}^{}_{\rm MNSP}&=& 
\mathcal{R}_1(\theta_\oplus) \mathcal{R}_2(\theta_{13},\delta_{\rm MNSP}) 
\mathcal{R}_3(\theta_\odot)\mathcal{P}\\
&\equiv& 
\pmatrix{1&0&0\cr
0&\cos \theta_{\oplus}&\sin\theta_{\oplus} \cr 0&-\sin\theta_{\oplus}
&\cos\theta_{\oplus}} \pmatrix{\cos\theta_{13} 
&0&\sin\theta_{13}e^{-i\delta_{\rm MNSP}}\cr
0&1&0\cr -\sin\theta_{13}e^{i\delta_{\rm MNSP}}&0&\cos 
\theta_{13}}\pmatrix{\cos\theta_{\odot} 
&\sin\theta_{\odot}&0\cr
-\sin\theta_{\odot}&\cos \theta_{\odot}&0\cr 0&0&1}\mathcal{P},\nonumber 
\eea
in which 
$\mathcal{R}_i(\theta)$ denote rotations by $\theta$ about the 
$i$th axis, and $\delta_{\rm MNSP}$ and the diagonal matrix $\mathcal{P}$ 
are CP-violating phases ($\mathcal{P}$ is physical only if 
neutrinos are Majorana fermions \cite{Bilenky:1980cx}). 
Combined 
data from solar 
(SNO-salt \cite{salt,Aharmim:2005gt} and Super-Kamiokande \cite{sksolar}), 
atmospheric \cite{skatm}, reactor \cite{kamland,chooz}, 
and accelerator \cite{k2k} experiments yield at 3$\sigma$
\cite{global}:
\be \quad \theta^{}_\oplus~=~ 
{45^\circ_{}}^{\,+10^\circ}_{\,-10^\circ} \ ;\quad 
\theta_{\odot}={33.9^\circ}^{+2.4^\circ}_{-2.3^\circ} \ ; \quad 
\theta_{13}< ~13^\circ_{}. 
\label{data1} 
\ee 
The rather precise constraints on the solar angle are indicated by the 
recent SNO data \cite{Aharmim:2005gt} when combined with KamLAND results 
\cite{kamland}. The atmospheric angle is consistent with maximal mixing, 
while maximal solar mixing is ruled out.  Presently there are no 
experimental constraints on the CP-violating phases. Note that the 
limiting value of the CHOOZ angle $\theta_{13}$ is approximately equal to 
$\theta_c$.

The CKM quark mixing matrix in the standard 
parametrization is 
\be 
\mathcal{U}^{}_{\rm CKM}=\mathcal{R}_1(\theta^{\rm CKM}_{23}) 
\mathcal{R}_2(\theta^{\rm CKM}_{13},\delta_{\rm CKM}) 
\mathcal{R}_3(\theta^{\rm CKM}_{12}) 
\ee 
has three small mixing angles 
and one ${\cal O}(1)$ CP-violating phase \cite{pdg}: 
\be 
\quad \theta^{\rm 
CKM}_{12} ~=~ {13.0^\circ_{}}\pm 0.1^\circ_{} \simeq \theta_c \ ;\quad 
\theta^{\rm CKM}_{23}=2.4^\circ_{}\pm 0.1^\circ_{}\ ;\quad \theta^{\rm 
CKM}_{13}=0.2^\circ_{}\pm 0.1^\circ_{}\ ;\quad \delta_{\rm 
CKM}=60^\circ\pm 14^\circ. 
\label{data3} 
\ee 
The CKM matrix is therefore approximately the identity matrix up to 
effects of order $\lambda$, as encoded by the Wolfenstein parametrization 
\cite{wolfenstein}:
\be {\cal U}_{\rm 
CKM}=\pmatrix{1-\frac{\lambda^2}{2}&\lambda&A\lambda^3(\rho-i\eta) 
\cr -\lambda &1-\frac{\lambda^2}{2}&A\lambda^2 \cr 
A\lambda^3(1-\rho-i\eta) &-A\lambda^2&1}+\mathcal{O}(\lambda^4), 
\label{wolf}
\ee 
in which $\lambda$ and $A$ are well known ($\lambda=0.22$, 
$A\simeq0.85$), but $\rho$ and $\eta$ are less precisely determined, 
reflecting the uncertainty in $\delta_{\rm CKM}$ \cite{pdg}.  
Although $\delta_{\rm CKM}$ is large (the only large angle in 
the quark sector), the 
rephasing invariant measure of CP violation (JDGW invariant)
\cite{jarlskog,greenberg}, which is given by 
\be \mathcal{J}_{\rm CP}
={\rm 
Im}(\mathcal{U}_{\alpha i}\mathcal{U}_{\beta j} \mathcal{U}^*_{\beta i} 
\mathcal{U}^*_{\alpha j}), 
\label{jarlskogdef} 
\ee is suppressed by small mixing angles: 
\be J^{\rm (CKM)}_{\rm CP}\simeq \sin 2\theta^{\rm CKM}_{12}\sin 
2\theta^{\rm CKM}_{23}\sin 2\theta^{\rm CKM}_{13} \sin\delta_{\rm 
CKM}\simeq 
A^2\lambda^6\eta \sim \mathcal{O}(10^{-5}). 
\ee 

From the perspective of quark-lepton unification, the discrepancy between 
the quark and lepton mixings provides an interesting challenge for 
theories of flavor.  A standard approach to flavor model building 
(see e.g. \cite{Altarelli:2004za} for a review) within 
this rather general framework is to build top-down models of the 
fermion mass matrices based on flavor symmetries which are typically 
broken at high scales, with the order parameter of the symmetry breaking 
set by $\lambda$.  These models may or may not be restricted to a 
particular grand unified theory (GUT).  The suppression of the 
neutrino masses can be linked with such high scales via the seesaw 
mechanism \cite{SEESAW}.

While perhaps surprising, the qualitative differences between the quark 
and lepton mixings are not inconsistent with grand unification.  The 
seesaw mechanism introduces a new unitary matrix into the MNSP matrix that 
has no analogue in the CKM, which can provide a source for the 
discrepancy. To see this more clearly, recall that the mixing matrices are 
the product of left-handed rotations for fermions of 
charge $q$ as follows: 
\bea {\cal U}^{}_{\rm CKM}~&=&~{\cal U}^{\dagger}_{2/3}\,{\cal 
U}^{}_{-1/3}\nonumber \\ 
{\cal U}^{}_{\rm MNSP}~&=&~{\cal U}^{\dagger}_{-1}\,\widehat{{\cal 
U}}^{}_{0}. 
\eea 
The quark and charged lepton rotation matrices $\mathcal{U}_q$ are 
obtained from diagonalizing their associated Dirac mass matrices. The form 
of the neutrino rotation matrix depends on whether neutrinos are 
Dirac or Majorana particles.  For Dirac neutrinos, 
$\widehat{\mathcal{U}}_0=\mathcal{U}_0$, while for the theoretically 
well-motivated case of the neutrino seesaw, 
$\widehat{\mathcal{U}}_0=\mathcal{U}_0\mathcal{F}$ \cite{datta,ramond}.  
Hence, 
for seesaw models the MNSP matrix takes the form
\be 
{\cal U}^{}_{\rm MNSP}~=~{\cal U}^{\dagger}_{-1}\,{\cal U}^{}_{0}\,\mathcal{F}, 
\label{ufumnsp} 
\ee 
which highlights the difference betweeen the quark and lepton mixing 
matrices. Models of the MNSP matrix can then be classified 
both by the structure of the lepton Dirac mass matrices and the 
number of large angles in $\mathcal F$ as given by the structure of the 
neutrino seesaw.

Within grand unification, possible connections can exist 
between the MNSP and CKM matrices.  In one illustrative class 
of examples, the fermion Dirac mass matrices 
$\mathcal{M}^{(q)}$ obey $SU(5)$ and $SO(10)$ GUT relations based on the 
simplest Higgs structures and the down quark mass matrix 
is further assumed to be symmetric, such that 
$\mathcal{M}^{(-1/3)}= \mathcal{M}^{(-1/3)\,T}\sim \mathcal{M}^{(-1)}$ 
and $\mathcal{M}^{(2/3)}\sim \mathcal{M}^{(0)}$.
In this case, the MNSP and CKM matrices 
are simply related \cite{ramond}: 
\be {\mathcal U}^{}_{\rm MNSP}~=~ {\mathcal U}^{\dagger}_{\rm 
CKM}\,\,{\mathcal F}\ 
\label{udagf}. 
\ee 
$\mathcal F$ must then contain two large mixing angles $\eta_{\odot}$ and 
$\eta_{\oplus}$. Eq.~(\ref{udagf}) implies that 
$\theta_{\odot}\sim 
\eta_{\odot} \pm \lambda \cos\eta_{\oplus}$ and $\theta_{13}\sim 
{\lambda}\sin\eta_{\oplus}$ due to the ${\cal O}(\lambda)$ mixing between 
the first and second families in ${\cal U}_{\rm CKM}$.

Another class of examples is motivated by the empirical fact that the 
measured values of the atmospheric and solar angles differ by 
$\sim \theta_c$, a relation known as quark-lepton complementarity 
\cite{Minakata:2004xt,
Smirnov:2004ju,
Raidal:2004iw,
Petcov:2004rk,
Ferrandis:2004mq,
Ferrandis:2004vp,
Ohlsson:2002rb,
Giunti:2002ye,Rodejohann:2003sc,
Frampton:2004ud,
Frampton:2004vw,Cheung:2005gq,
Kang:2005as,Antusch:2005ca,Ohlsson:2005js,Antusch:2005kw}. 
$\mathcal{U}_{\rm MNSP}$ then initially can be bimaximal 
($\eta_{\odot}=\eta_{\oplus}=45^{\circ}$), with the solar angle shifted by 
a full-strength Cabibbo shift: $\theta_{\odot}\sim \eta_{\odot} - 
\theta_c$.  $\theta_{13}$ can also be a Cabibbo effect (see e.g. 
\cite{Antusch:2005ca}).

While these examples can be motivated by flavor theories, one should keep 
in mind that the experimental measurements are not yet precise enough to 
select particular 
scenarios and since the values of the lepton mixing angles are cloaked 
from us by the Cabibbo haze (if indeed the $\lambda\rightarrow 0$ limit is 
meaningful for theory\footnote{An alternative expansion based on rational 
hierarchy and $S3-S2$ symmetry with an expansion parameter similar in 
spirit (though not in magnitude) to the Cabibbo angle can be found in 
\cite{Kaus:2004ix}.}), many possibilities exist beyond these prototype 
examples (see e.g. \cite{Antusch:2005kw}). A taxonomy of parametrizations 
of the MNSP matrix for  
models in which the bare CHOOZ angle vanishes was carried 
out in \cite{Datta:2005ci}.  We now develop and extend 
this formalism to additional models based on the physics of lepton mixing in 
the $\lambda\rightarrow 0$ limit.
   
\section{Theoretical Framework} 
The Wolfenstein parametrization is based on the idea that the hierarchical 
quark mixing angles can be understood as a $\lambda$ expansion, with 
\be 
\mathcal{U}_{\rm CKM}=1+\mathcal{O}(\lambda). 
\ee 
In the lepton sector, a similar parametrization for the MNSP matrix 
requires a $\lambda$ expansion of the form
\be 
\mathcal{U}_{\rm MNSP}=\mathcal{W}+\mathcal{O}(\lambda). 
\ee 
The starting matrix $\mathcal{W}$, which is dictated by the 
(unknown) underlying flavor theory, is then perturbed multiplicatively by a 
unitary matrix $\mathcal{V}(\lambda)$, which in turn is assumed to have a 
$\lambda$ expansion:
\be
\mathcal{V}(\lambda)=1+\mathcal{O}(\lambda).
\ee
For the quarks, the starting matrix is the identity matrix and the 
perturbation matrix is given by Eq.~(\ref{wolf}). For the leptons, the 
structure of the allowed perturbations depend on the details of 
$\mathcal{W}$.  Due to Cabibbo haze, $\mathcal{W}$ can take different 
forms which are characterized by the number of large angles.  
Neglecting Dirac CP-violating phases (they will be discussed later), the possibilities are: 
\begin{itemize} 
\item {\it Three large angles}.  In this scenario, $\mathcal{W}$ is given 
by
\bea 
\mathcal{W}&=&\mathcal{R}_1(\eta_\oplus)\mathcal{R}_2(\eta_{13}) 
\mathcal{R}_3(\eta_\odot)\mathcal{P}\nonumber \\
&\equiv&\pmatrix{1&0&0\cr
0&\cos \eta_{\oplus}&\sin\eta_{\oplus} \cr 0&-\sin\eta_{\oplus}
&\cos\eta_{\oplus}} \pmatrix{\cos\eta_{13} &0&\sin\eta_{13}\cr
0&1&0\cr -\sin\eta_{13}&0&\cos \eta_{13}}\pmatrix{\cos\eta_{\odot} 
&\sin\eta_{\odot}&0\cr
-\sin\eta_{\odot}&\cos \eta_{\odot}&0\cr 0&0&1}\mathcal{P}, 
\label{wdef3angle} 
\eea 
in which the angles 
$\eta_{\oplus}$, $\eta_{13}$, and $\eta_{\odot}$ correspond to the bare 
values of the atmospheric, CHOOZ, and solar angles, respectively.  
$\mathcal{P}$ is a diagonal phase matrix of the form 
\be 
\mathcal{P}= \pmatrix{e^{i\alpha_1}&0&0\cr 0&e^{i\alpha_2} &0\cr 
0&0&e^{i\alpha_3}}, 
\ee 
which encodes the two physical Majorana CP-violating phases 
$\alpha_{12}\equiv\alpha_1-\alpha_2$ and 
$\alpha_{23}\equiv\alpha_2-\alpha_3$. Since the CHOOZ angle is 
$\simlt \theta_c$, the Cabibbo-sized perturbations must downshift 
$\eta_{13}$ into the allowed range.

\item {\it Two large angles}.  
In this case, $\mathcal{W}$ contains two large angles 
$\eta_{\oplus}$, $\eta_{\odot}$ and a zero CHOOZ angle: 
\be 
{\cal W}={\cal R}_1(\eta_{\oplus}){\cal R}_{3}(\eta_{\odot})\equiv 
\pmatrix{1&0&0\cr 
0&\cos \eta_{\oplus}&\sin\eta_{\oplus} \cr 0&-\sin\eta_{\oplus} 
&\cos\eta_{\oplus}} \,\pmatrix{\cos\eta_{\odot} &\sin\eta_{\odot}&0\cr 
-\sin\eta_{\odot}&\cos \eta_{\odot}&0\cr 0&0&1}\mathcal{P}, 
\label{wdef2angle} 
\ee
which of course follows from Eq.~(\ref{wdef3angle}) with $\eta_{13}=0$.  
In this scenario, which was studied in detail in \cite{Datta:2005ci}, both 
the CHOOZ angle and the JDGW invariant are generated from Cabibbo shifts.
 
\item {\it One large angle}.  $\mathcal{W}$ takes the form
\be
{\cal W}={\cal R}_1(\eta_{\oplus})\mathcal{P}\equiv
\pmatrix{1&0&0\cr
0&\cos \eta_{\oplus}&\sin\eta_{\oplus} \cr 0&-\sin\eta_{\oplus}
&\cos\eta_{\oplus}}\mathcal{P}.
\ee
The solar and CHOOZ angles (and the JDGW invariant) arise from Cabibbo 
effects. Given the typical size of Cabibbo shifts, shifting the 
solar angle into the allowed range requires a large 
effect (e.g. a sum of several shifts).
Note that only one Majorana phase $\alpha_{23}$ is 
observable in the absence of the perturbations.

\item {\it No large angles}. In this case, all three mixing angles are 
generated by Cabibbo shifts.  Given the typical size of Cabibbo effects, 
it is unlikely that they are the dominant source of the atmospheric angle, 
for which the best-fit value is consistent with maximal mixing (as we will 
see, even shifting the solar angle in this way requires a stretching of 
parameters).  Therefore, we will not consider this case further in this paper.
\end{itemize}

Let us pause here to comment on Dirac phases, which in general may be 
present in $\mathcal{W}$.  If the freedom in the SM to rephase the charged 
leptons is taken into account, these Dirac phases can be rotated 
away if one or more of the bare mixing angles is zero since the JDGW 
invariant identically vanishes.
If $\mathcal{W}$ has three large angles, there is one physical combination 
of phases $\chi$ which enters in the 
standard parametrization through $\mathcal{R}_2(\eta_{13})\rightarrow 
\mathcal{R}_2(\eta_{13},\chi)$, in direct analogy with $\delta_{\rm MNSP}$ 
in Eq.~(\ref{umnspstandardparam}).  However, when determining 
$\mathcal{U}_{\rm MNSP}$ at higher orders in $\lambda$ it is too 
preliminary to do this rephasing at the $\lambda=0$ level, as the bare 
Dirac phases can enter physical observables at 
higher orders in $\lambda$.  For simplicity, we assume in this 
paper that these phases are absent, as otherwise the number of bare 
parameters proliferates.  In doing so, we assume that Dirac CP 
violation is necessarily linked to Cabibbo shifts.  This encompasses 
the main physics of models for which one or more of the bare mixing angles 
vanish. However, it is an extra assumption for models with three large 
initial mixing angles, placing constraints on the as yet unspecified 
underlying theory (which 
must then predict $\chi\rightarrow 0$).

A novel feature of these parametrizations of lepton mixing is 
that generically the perturbations do not commute with the starting 
matrix 
\be 
[{\mathcal W}\,,\,{{\mathcal V}(\lambda)}\,]~\neq~0\ , 
\ee 
leading to several possible implementations of the Cabibbo shifts:  
\begin{itemize} 
\item {\it Right Cabibbo Shifts}. The perturbations can be introduced as a 
multiplication of $\mathcal{V}(\lambda)$ on the right:
\be 
{\cal U}_{\rm MNSP}={\cal W}\,{\cal V}(\lambda). 
\label{eqright}
\ee

\item {\it Left Cabibbo Shifts}.  The perturbations can be 
implemented as a multiplication of  $\mathcal{V}(\lambda)$ on the left: 
\be 
{\cal U}_{\rm MNSP}={\cal V}(\lambda)\,{\cal W}. 
\label{eqleft}
\ee 

\item {\it Middle Cabibbo Shifts}.
The perturbations can be sandwiched between the rotation matrices of 
$\mathcal{W}$:
\be {\cal U}_{\rm MNSP}={\cal R}_{1}\,{\cal V(\lambda)}\,{\cal 
R}_{2}\,{\cal R}_3\mathcal{P}
\label{eqmidleft} 
\ee 
or
\be {\cal U}_{\rm MNSP}={\cal R}_{1}\,\mathcal{R}_2\,{\cal 
V}(\lambda)\,{\cal R}_3\mathcal{P}.
\label{eqmidright}
\ee
\end{itemize} 
If $\mathcal{W}$ has two large angles ($\eta_{13}\rightarrow 
0$), Eqs.~(\ref{eqmidleft}) and (\ref{eqmidright}) are equivalent.  For 
one large angle in $\mathcal{W}$ ($\eta_{13},\eta_\odot\rightarrow 
0$), the middle and right Cabibbo shifts are redundant. $\mathcal{V}$ 
can also be sandwiched between $\mathcal{R}_3$ and $\mathcal{P}$; 
aside from the effects of the Majorana phases (discussed 
later), this case is equivalent to right shifts for $\mathcal{P}=1$.    

At this stage, we pause to comment on the meaning of this 
classification of models into right, left, or 
middle Cabibbo shift scenarios.  Since this classification depends on the 
initial parametrization of $\mathcal{W}$, it is perhaps not clear whether 
it encompasses all possibilities.  To see that there is no loss of 
generality, recall that the assumption of Cabibbo haze is that the MNSP 
matrix has a $\lambda$ expansion 
\be
\mathcal{U}_{\rm 
MNSP}(\lambda)=\sum^{\infty}_{n=0}\lambda^nW_n,
\ee
in which $W_0\equiv\mathcal{W}$. There is a choice in how to parametrize 
the Cabibbo haze; for example, it can be expressed as a right 
Cabibbo shift as follows:
\be
\mathcal{U}_{\rm
MNSP}(\lambda)=\mathcal{W} 
\sum^{\infty}_{n=0}\lambda^n(\mathcal{W}^{-1}W_n) 
\equiv\mathcal{W}\mathcal{V}(\lambda), 
\ee
identifying 
$\mathcal{V}=\sum^{\infty}_{n=0}\lambda^n(\mathcal{W}^{-1}W_n)$.
However, recalling that $\mathcal{W}$ is given by a product 
of rotation matrices $\mathcal{R}_i$ 
(neglecting $\mathcal{P}$ at the moment for simplicity, though it is 
straightforward to include it), this can also be expressed as a middle 
Cabibbo shift:
\bea
\mathcal{U}_{\rm 
MNSP}(\lambda)&=&\mathcal{R}_1\mathcal{R}_2\mathcal{R}_3 
\mathcal{V}(\lambda)\nonumber \\  
&=&\mathcal{R}_1\mathcal{R}_2\left (\mathcal{R}_3
\mathcal{V}(\lambda)\mathcal{R}^{-1}_3 \right 
)\mathcal{R}_3\equiv \mathcal{R}_1\mathcal{R}_2 
\mathcal{V}^{\prime}(\lambda)\mathcal{R}_3. \eea
The generalization to other middle shifts and to left shifts is 
straightforward.  Note that since $\mathcal{V}$ by assumption represents 
perturbations about the identity matrix
\be
\mathcal{V}(\lambda)=1+\sum^{\infty}_{i=1}\lambda^nV_n,
\ee
$\mathcal{V}^\prime$ can also be written in an analogous form:
\be
\mathcal{V}^\prime(\lambda)=\mathcal{R}_3\mathcal{V}\mathcal{R}^{-1}_3 
=1+\sum^{\infty}_{i=1}\lambda^n(\mathcal{R}_3V_n\mathcal{R}^{-1}_3).
\ee
Therefore, the 
decomposition into right, left, or middle shifts is meaningful for any 
{\it specific} choice of $\mathcal{V}$ (which may be illuminating in the 
context of the underlying flavor theory).  
For the purposes of our classification scheme, we will assume 
$\mathcal{V}$ takes a simple form.  More precisely, we write 
$\mathcal{V}=e^{\mathcal{A}}$, with 
\be
\mathcal{A}=\pmatrix{0 &\,\,\,\,\,\, \sum_{i=1} a_i\lambda^i& \sum_{i=1} 
c_i\lambda^i\cr
-\sum_{i=1} a^*_i\lambda^i&\,\,\,\,\,\, 0& \sum_{i=1} b_i\lambda^i\cr 
-\sum_{i=1} 
c^*_i\lambda^i&-\sum_{i=1} b^*_i\lambda^i&0},
\label{aexpr}
\ee
in which $a_i$, $b_i$, and $c_i$ are ${\cal O}(1)$ 
coefficients.\footnote{We point out a change in notation here from 
\cite{Datta:2005ci}, which used $(a,\,a',\,a'')$ to denote coefficients 
of ${\cal O}(\lambda)$ terms, $(b,\,b',\,b'')$ for ${\cal O}(\lambda^2)$ 
terms, and $(c,\,c',\,c'')$ for ${\cal O}(\lambda^3)$ terms.}  
$\mathcal{V}(\lambda)$ is 
written as an 
expansion in  $\lambda$, such that e.g. through ${\cal O}(\lambda^2)$, 
\be
\mathcal{V}=\pmatrix{1-\frac{|a_1|^2+|c_1|^2}{2}\lambda^2 & 
a_1\lambda+(a_2-\frac{b^*_1c_1}{2})\lambda^2&  
c_1\lambda+(c_2+\frac{a_1b_1}{2})\lambda^2\cr
-a^*_1\lambda-(a^*_2-\frac{b_1c^*_1}{2})\lambda^2&
1-\frac{|a_1|^2+|b_1|^2}{2}\lambda^2 & 
b_1\lambda+(b_2-\frac{a^*_1c_1}{2})\lambda^2\cr 
-c^*_1\lambda-(c^*_2-\frac{a^*_1b^*_1}{2})\lambda^2
& -b^*_1\lambda-(b^*_2+\frac{a_1c^*_1}{2})\lambda^2&
1-\frac{|b_1|^2+|c_1|^2}{2}\lambda^2}+\mathcal{O}(\lambda^3).
\label{vtolamsq}
\ee
This expression for $\mathcal{V}$ encodes the Wolfenstein 
form of the CKM, which can be obtained by choosing $a_1=1$, $b_2=A$, 
$c_3=A (\rho-\frac{1}{2}-i\eta)$, and $b_1=c_{1,2}=0$, but also 
allows for more general perturbations.\footnote{Note that 
Eq.~(\ref{aexpr}) does not lead to the most general unitary matrix, 
because $\mathcal{A}$ can have diagonal entries which are purely 
imaginary, which yields additional phases in $\mathcal{V}$ than what 
appears in Eq.~(\ref{vtolamsq}). These 
phases can be removed by global rephasings; for example, this is what is 
done for the CKM, which has only one observable phase.  
Although there is no {\it a priori} reason why such additional phase 
degrees of freedom should not be present in $\mathcal{V}$,
for the sake of simplicity we will not include them in this paper.}

The shifts in the mixing angles induced by $\mathcal{V}(\lambda)$ are 
clearly dominated by perturbations linear in $\lambda$, 
which lead to shifts of at most $\sim \theta_c\simeq13^\circ$. The 
${\cal O}(\lambda^2)$ perturbations, which lead to shifts of $\sim 3^\circ$ or 
so, play a subdominant role but can be important in certain models.  
Therefore, it is useful to categorize 
models further according to the number of ${\cal O}(\lambda)$ 
perturbations \cite{Datta:2005ci}:
\begin{itemize}
\item {\it Single shifts.} These models have one ${\cal O}(\lambda)$ 
perturbation, {\it i.e.} either (i) $a_1\neq 0$ and $b_1=c_1=0$, (ii) 
$b_1\neq 0$ and $a_1=c_1=0$, or (iii) $c_1\neq 0$ and $a_1=b_1=0$.

\item {\it Double shifts.} These models contain two ${\cal O}(\lambda)$ 
perturbations, again with three possibilities: (i) $a_1,\,b_1\neq 0$ 
and $c_1=0$, (ii) $b_1,\,c_1\neq 0$ and $a_1=0$, or (iii) $a_1,\,c_1\neq 
0$ and $b_1=0$.

\item {\it Triple shifts.}  In these models, $a_1$, $b_1$, and $c_1$ are 
all nonvanishing.
\end{itemize}
Double and triple shifts can also be implemented in more 
complicated ways.  One possibility is to assume sequential perturbations 
of the form $\mathcal{V}=\mathcal{V}_1\mathcal{V}_2$, in which 
$\mathcal{V}_1$ is given by Eq.~(\ref{vtolamsq}), and $\mathcal{V}_2$ is 
given by the same expression after taking $\{a_i,b_i,c_i\}\rightarrow 
\{a'_i,b'_i,c'_i\}$. In this case, double shift models can also be 
achieved when $\mathcal{V}_1$ and $\mathcal{V}_2$ are both given by single 
shifts (and a similar argument holds for triple shifts). Additional middle 
(or ``mixed") Cabibbo shift scenarios can then be constructed by 
sandwiching $\mathcal{V}_1$ and $\mathcal{V}_2$ among the rotation matrices of 
$\mathcal{W}$.  Such methods of incorporating double and triple shifts 
were mentioned in \cite{Datta:2005ci}, but they add little to the 
qualitative conclusions for models in which the required shifts in the 
lepton mixing angles are of a natural size ($\sim \theta_c$). 
However, given that we consider scenarios which require 
larger shifts, we will keep such sequential perturbations in mind 
in this paper because they allow for larger effects if $\mathcal{V}_{1,2}$ 
are both given by double or triple shifts (though of course, at 
leading order in $\lambda$ sequential perturbations are equivalent to a 
``standard" perturbation with unnaturally large coefficients).

For all models, an ${\cal O}(\lambda)$ 
entry in $\mathcal{V}(\lambda)$ gives rise to an ${\cal O}(\lambda)$ 
Cabibbo shift in the corresponding mixing angle.  For 
example, a nonvanishing $a_1$ shifts the solar angle by a Cabibbo-sized 
effect (which may be sized by factors which depend on the bare MNSP 
parameters).  Shifts can also be incurred from the 
$\mathcal{O}(\lambda)$ perturbations in other entries of 
$\mathcal{V}(\lambda)$; these shifts depend on the structure of 
$\mathcal{W}$ and the ways in which the Cabibbo shifts are introduced.  
For example, for the case of two large initial angles \cite{Datta:2005ci}, 
single shift models with nonvanishing 
$b_1$ shift both the atmsopheric and CHOOZ angles for right shifts, but 
shift the atmospheric angle only for left and middle shifts.  One of the 
purposes of this paper is to analyze the structure of the Cabibbo shifts 
for the scenarios with other choices of $\mathcal{W}$, an issue to which 
we now turn.

\section{Parametrizations} 
In this section, we discuss the leptonic Cabibbo shifts for the 
Wolfenstein-like parametrizations of the MNSP matrix.  We 
first present the shifts in the MNSP mixing angles to 
$\mathcal{O}(\lambda)$, assuming 
$\mathcal{V}(\lambda)$ is given by Eq.~(\ref{vtolamsq}) (unless otherwise 
specified).  The discussion of CP violation is deferred to the next 
subsection.  We assume bare Dirac phases are absent, but 
include bare Majorana phases and phases in $\mathcal{V}(\lambda)$.

Before discussing specific models, it is important to keep in mind that 
there is a wide range of possible parametrizations that can be 
constructed by specifying the details of $\mathcal{W}$ 
and $\mathcal{V}(\lambda)$.  While the leading order perturbations dictate the 
dominant Cabibbo shifts, subleading terms can also play an important 
role (particularly for the 
CP-violating effects, as discussed later). 
There is also a further range of possibilities for each of the 
given scenarios due to the experimental uncertainties in the MNSP parameters.  
This is particularly at play for the atmospheric and CHOOZ angles, which 
have experimental uncertainties which are roughly of ${\cal O}(\lambda)$. 
The solar angle is measured more precisely, with error bars of ${\cal 
O}(\lambda^2)$.  

Given the wide range of scenarios consistent with current data,  
we choose not to analyze specific numerical examples at this stage. 
Particular parametrizations may emerge as compelling from the 
standpoint of flavor theory, warranting further analysis.  
Improved data will be invaluable in narrowing the 
range of possible parametrizations. Although improvement in the 
atmospheric angle is not expected in the foreseeable future \cite{kajita}, 
the planned reactor neutrino experiments, superbeams and/or neutrino factories 
are expected to probe the CHOOZ angle from its current ${\cal O}(\lambda)$ 
range down to the ${\cal O}(\lambda^2)\sim 3^\circ$ level \cite{nufact}. 
It is therefore useful to classify scenarios according to whether they 
predict a CHOOZ angle of ${\cal O}(\lambda)$ or further suppressed.  
Hence, we now present a complete catalog of the ${\cal 
O}(\lambda)$ leptonic Cabibbo shifts for each scenario of interest.

\subsection{Cabibbo-shifted lepton mixing angles}
\subsubsection*{Two large angles.}
We begin by reviewing and summarizing the results for two large 
initial mixing angles.  Given the patterns of the data as well as the 
philosophy of treating the Cabibbo angle as a small expansion parameter, 
this scenario is arguably the most plausible starting point, and has been 
the primary focus for most flavor model building attempts.  
For this reason, this class of models was first discussed as a prototype example of 
Cabibbo haze in \cite{Datta:2005ci}, to which we refer the reader for 
more details.  Here we note for reference the following general results 
for the ${\cal O}(\lambda)$ shifts in the mixing angles (including 
phases):\\

\noindent $\bullet$ {\it Right shifts}: $\mathcal{U}_{\rm 
MNSP}=\mathcal{R}_1(\eta_\oplus)\mathcal{R}_3(\eta_\odot)\mathcal{P} 
\mathcal{V}(\lambda)$. 
\bea
\label{2rcssol}
\theta_\odot&=&\eta_\odot+\lambda |a_1|\cos(\alpha_{12}+\phi_{a_1}) 
+\mathcal{O}(\lambda^2)\\
\label{2rcsatm}
\theta_\oplus&=&\eta_\oplus+\lambda 
(\cos\eta_\odot|b_1|\cos(\alpha_{23}+\phi_{b_1})
-\sin\eta_\odot|c_1|\cos(\alpha_{12}-\alpha_{23}+\phi_{c_1})) 
+\mathcal{O}(\lambda^2)\\
\label{2rcschooz}
\theta_{13}&=&\lambda|b_1e^{i\alpha_{23}}\sin\eta_\odot
+c_1e^{i(\alpha_{12}-\alpha_{23})}\cos\eta_\odot| 
+\mathcal{O}(\lambda^2). 
\eea

\noindent $\bullet$ {\it Left shifts}: $\mathcal{U}_{\rm
MNSP}=\mathcal{V}(\lambda)
\mathcal{R}_1(\eta_\oplus)\mathcal{R}_3(\eta_\odot)\mathcal{P}$.
\bea
\label{2lcssol}
\theta_\odot&=&\eta_\odot+\lambda(\cos\eta_\oplus 
|a_1|\cos\phi_{a_1}-\sin\eta_\oplus|c_1|\cos\phi_{c_1}) 
+\mathcal{O}(\lambda^2)\\
\label{2lcsatm}
\theta_\oplus&=&\eta_\oplus+\lambda|b_1|\cos\phi_{b_1} 
+\mathcal{O}(\lambda^2)\\
\label{2lcschooz}
\theta_{13}&=&\lambda|\sin\eta_{\oplus}a_1 + 
\cos\eta_\oplus c_1|+\mathcal{O}(\lambda^2).
\eea

\noindent $\bullet$ {\it Middle shifts}:  $\mathcal{U}_{\rm
MNSP}=\mathcal{R}_1(\eta_\oplus)\mathcal{V}(\lambda)
\mathcal{R}_3(\eta_\odot)\mathcal{P}$.
\bea
\label{2mcssol}
\theta_\odot&=&\eta_\odot+\lambda|a_1|\cos\phi_{a_1} 
+\mathcal{O}(\lambda^2)\\
\label{2mcsatm}
\theta_\oplus&=&\eta_\oplus+\lambda|b_1|\cos\phi_{b_1} 
+\mathcal{O}(\lambda^2)\\
\label{2mcschooz}
\theta_{13}&=&\lambda|c_1|+\mathcal{O}(\lambda^2).
\eea
These results demonstrate that the shifts in the mixing angles depend on 
the Majorana phases $\alpha_{12},\,\alpha_{23}$ only in the right 
Cabibbo shift scenario.  
The special status of Majorana phases for right Cabibbo shifts is a 
general theme that will continue throughout this paper, and is a result 
which is easy to understand.  More precisely, the presence of Majorana 
phases in $\mathcal{U}_{\rm MNSP}$ is due to the lack of 
freedom to rephase Majorana fermions, a feature which is encoded 
(in the standard parametrization) as a diagonal phase matrix 
multiplied on the far right.  Since right shifts introduce the 
perturbations $\mathcal{V}(\lambda)$ in that position and generically
\be
[{\mathcal P}\,,\,{{\mathcal V}(\lambda)}\,]\neq 0, 
\ee
the right shifts can be rewritten as follows:
\bea
\label{modrightshifts}
\mathcal{U}_{\rm MNSP}&=&\mathcal{W}\mathcal{P}\mathcal{V}\nonumber \\
&=&\mathcal{W}(\mathcal{P}\mathcal{V}\mathcal{P}^{-1})\mathcal{P} \equiv 
\mathcal{W}\mathcal{V}_{\mathcal{M}}\mathcal{P}.
\eea
Eq.~(\ref{modrightshifts}) implies that $\mathcal{V}_{\mathcal{M}}$ can 
be obtained from $\mathcal{V}$ through the replacements $a_i\rightarrow 
a_ie^{i\alpha_{12}}$, $b_i\rightarrow b_ie^{i\alpha_{23}}$, and 
$c_i\rightarrow c_ie^{i(\alpha_{12}-\alpha_{23})}$, a result which is 
manifest in Eqs.~(\ref{2rcssol})--(\ref{2rcschooz}). 

Before moving on, we stress once again that for a specific form of 
$\mathcal{V}$, different shift scenarios can lead to very different 
results for the mixing angles.  For example, if one assumes that 
$\mathcal{V}$ has the same hierarchical structure at leading order as the 
CKM ({\it i.e.}, $b_1=c_1=0$, $a_1\neq 0$), models with a bimaximal 
starting matrix $\mathcal{W}$ (as motivated e.g. by quark-lepton 
complementarity) can in principle be consistent with the data for right 
shifts and middle shifts; note that in both cases the CHOOZ angle is 
predicted to be subleading in $\lambda$.  However, the left shift scenario 
does not fit the data in this case, as the shift in the 
solar angle is sized by factors dependent on the bare angles (amounting 
here to a suppression).  Of course, within our approach there are other 
ways to make this starting matrix consistent with the 
data, as there is no reason {\it a priori} to focus only on perturbations 
which resemble the CKM.  For example, 
left shifts can also work if in addition to ${\cal O}(\lambda)$ mixing 
between the first and second generations, the 
perturbations have ${\cal O}(\lambda)$ mixing between the first and 
third families (see Eqs.~(\ref{2lcssol})--(\ref{2lcschooz})).

\subsubsection*{Three large angles: Cabibbo-downshifted CHOOZ angle.} 
We now turn to the most general scenario, which has three 
nonvanishing angles in $\mathcal{W}$.  This structure can be motivated 
e.g. by the idea of neutrino ``anarchy" first proposed in 
\cite{Hall:1999sn}, which 
predicts neutrino mass matrices with random 
${\cal O}(1)$ entries, leading to three large mixing angles. Although 
the current bound on the CHOOZ angle is small, it is not vanishingly 
small, and hence the anarchy hypothesis can provide a viable framework for 
flavor model building.  If neutrino anarchy can be consistently embedded 
within grand unification, Cabibbo-sized perturbations can act 
to reduce the CHOOZ angle from its typically large starting value, which 
could potentially open up new avenues for model building.

Given this form of $\mathcal{W}$ and $\mathcal{V}$ given by 
Eq.~(\ref{vtolamsq}), the mixing angles for the Cabibbo shift scenarios 
are: \\

\noindent $\bullet$ {\it Right shifts}:  $\mathcal{U}_{\rm
MNSP}=\mathcal{R}_1(\eta_\oplus)\mathcal{R}_2(\eta_{13}) 
\mathcal{R}_3(\eta_\odot)\mathcal{P}\mathcal{V}(\lambda)$.

\bea
\label{3rcssol}
\theta_\odot&=&\eta_\odot+\lambda (|a_1|\cos(\alpha_{12}+\phi_{a_1})-
\nonumber \\
& & 
\tan\eta_{13}(\cos\eta_\odot|b_1|\cos(\alpha_{23}+\phi_{b_1})
+ \sin\eta_\odot 
|c_1|\cos(\alpha_{12}-\alpha_{23}+\phi_{c_1})))+\mathcal{O}(\lambda^2)\\
\label{3rcsatm}
\theta_\oplus&=&\eta_\oplus+\frac{\lambda}{\cos\eta_{13}} 
(\cos\eta_\odot|b_1|\cos(\alpha_{23}+\phi_{b_1})
-\sin\eta_\odot|c_1|\cos(\alpha_{12}-\alpha_{23}+ 
\phi_{c_1}))+\mathcal{O}(\lambda^2)\\
\label{3rcschooz}
\theta_{13}&=&\eta_{13}+\lambda(\sin\eta_\odot|b_1|\cos(\alpha_{23}+\phi_{b_1})
+\cos\eta_\odot|c_1|\cos(\alpha_{12}-\alpha_{23}+\phi_{c_1}))+ 
\mathcal{O}(\lambda^2). 
\eea

\noindent $\bullet$ {\it Left shifts}: $\mathcal{U}_{\rm
MNSP}=\mathcal{V}(\lambda) 
\mathcal{R}_1(\eta_\oplus)\mathcal{R}_2(\eta_{13})
\mathcal{R}_3(\eta_\odot)\mathcal{P}$.
\bea
\label{3lcssol}
\theta_\odot&=&\eta_\odot+\frac{\lambda}{\cos\eta_{13}}(\cos\eta_\oplus 
|a_1|\cos\phi_{a_1}-\sin\eta_\oplus|c_1|\cos\phi_{c_1}) 
+\mathcal{O}(\lambda^2)\\
\label{3lcsatm}
\theta_\oplus&=&\eta_\oplus+\lambda(|b_1|\cos\phi_{b_1}-
\tan\eta_{13}(\cos\eta_\oplus|a_1|\cos\phi_{a_1}-
\sin\eta_\oplus|c_1|\cos\phi_{c_1}))+\mathcal{O}(\lambda^2)\\
\label{3lcschooz}
\theta_{13}&=&\eta_{13}+\lambda(\sin\eta_{\oplus}|a_1|\cos\phi_{a_1} + 
\cos\eta_\oplus |c_1|\cos\phi_{c_1})+\mathcal{O}(\lambda^2).
\eea

\noindent $\bullet$ {\it Middle left shifts}: $\mathcal{U}_{\rm
MNSP}=\mathcal{R}_1(\eta_\oplus)\mathcal{V}(\lambda)
\mathcal{R}_2(\eta_{13})\mathcal{R}_3(\eta_\odot)\mathcal{P}$.
\bea
\label{3mlcssol}
\theta_\odot&=&\eta_\odot+\frac{\lambda}{\cos\eta_{13}}|a_1|\cos\phi_{a_1} 
+\mathcal{O}(\lambda^2)\\
\label{3mlcsatm}
\theta_\oplus&=&\eta_\oplus+\lambda(|b_1|\cos\phi_{b_1}- 
\tan\eta_{13}|a_1|\cos\phi_{a_1}) +\mathcal{O}(\lambda^2)\\
\label{3mlcschooz}
\theta_{13}&=&\eta_{13}+\lambda|c_1|\cos\phi_{c_1} 
+\mathcal{O}(\lambda^2).
\eea

\noindent $\bullet$ {\it Middle right shifts}: $\mathcal{U}_{\rm
MNSP}=\mathcal{R}_1(\eta_\oplus)\mathcal{R}_2(\eta_{13})
\mathcal{V}(\lambda)
\mathcal{R}_3(\eta_\odot)\mathcal{P}$.
\bea
\label{3mrcssol}
\theta_\odot&=&\eta_\odot+\lambda(|a_1|\cos\phi_{a_1}- 
\tan\eta_{13}|b_1|\cos\phi_{b_1}) +\mathcal{O}(\lambda^2)\\
\label{3mrcsatm}
\theta_\oplus&=&\eta_\oplus+\frac{\lambda}{\cos\eta_{13}}|b_1|\cos\phi_{b_1} 
+\mathcal{O}(\lambda^2)\\
\label{3mrcschooz}
\theta_{13}&=&\eta_{13}+\lambda|c_1|\cos\phi_{c_1} 
+\mathcal{O}(\lambda^2).
\eea
As in the previous case, each Cabibbo shift scenario leads to a particular 
pattern of shifted mixing angles for a given $\mathcal{W}$ and 
$\mathcal{V}$.  Since in this scenario the primary issue is to downshift 
the CHOOZ angle, it is instructive to compare the correlations between the 
dominant shifts in $\theta_{13}$ and the other angles for each scenario.  
Right and left shifts imply particular correlations between the shift in 
$\theta_{13}$ and the shifts in $\theta_\odot$ and $\theta_\oplus$ (with 
the right shifts involving the Majorana phases as usual). In contrast, for 
the two middle shift scenarios, the dominant shift in the CHOOZ angle is 
uncorrelated with the dominant shifts in the other two mixing angles.

There are several ways to construct viable scenarios given a particular 
choice of the bare angles $\eta_\odot$, $\eta_\oplus$, and $\eta_{13}$.  
Although we defer a detailed analysis of specific 
scenarios for future study, consider as an illustrative example a 
scenario in which all three bare mixing angles are close to their maximal 
values.  In this case, both a sizeable 
downshift in the CHOOZ angle (of order $\simgt 2\theta_c$ or so)  and a 
downshift in the solar angle of $\sim\theta_c$
are required. Let us assume that with a suitable choice of parameters the 
CHOOZ angle can be shifted into the allowed range
(we will return to this issue shortly).  The question is 
then whether the solar angle can be shifted into the rather 
precise range allowed by the data.\footnote{Note that ${\cal 
O}(\lambda^2)$ terms should also be included since large shifts 
are required, although they are not displayed here.}

For right shifts, Eqs.~(\ref{3rcssol})--(\ref{3rcschooz}) show that 
downshifting $\theta_{13}$ and $\theta_\odot$ either implies 
particular shifts (e.g. in the atmospheric angle) or dictates the sizes of 
other coefficients (see Eq.~(\ref{3rcssol})). In addition, since this 
scenario has a tendency toward anticorrelation between the shifts of the 
solar and CHOOZ angles, in general the parameters must be stretched for 
the ratio of these two shifts to be within the ballpark of the data, 
independently of whether the large shift in the CHOOZ angle can be 
achieved without fine-tuning.  For left shifts, an inspection 
of Eqs.~(\ref{3lcssol})--(\ref{3lcschooz}) shows that the required ratio 
of the downshifts of the solar and CHOOZ angles is not as difficult to 
achieve.  For both middle shift scenarios the 
size of the dominant shift in the CHOOZ angle is unconnected with the 
shift in the solar angle, so independent coefficients govern the ratio of 
their shifts.

In this example, the bare CHOOZ angle is very different from its 
experimental value, requiring a large Cabibbo shift for consistency with 
the data.  Such large shifts can be difficult to achieve for perturbations 
of the form given in Eq.~(\ref{vtolamsq}) with ${\cal O}(1)$ coefficients.  
Obtaining large shifts can be facilitated by assuming sequential 
perturbations (as mentioned previously).  
Sequential perturbations also allow for many additional mixed Cabibbo 
shift scenarios, corresponding to the many ways to introduce these two 
perturbation matrices in the MNSP matrix.  We do not enumerate all the 
possibilities here, but rather comment that such scenarios may be of 
theoretical interest because they can more easily make a ``trimaximal" 
starting matrix consistent with the data.  We will 
return to this idea when discussing models in which the solar angle is 
sourced by Cabibbo effects.

Of course, trimaximal or other starting matrices which require a large 
shift in the CHOOZ angle are not the only option.  For perturbations of 
the usual type (Eq.~(\ref{vtolamsq}) without unnaturally large 
coefficients), it is much easier to accommodate scenarios in which the 
initial CHOOZ angle is perhaps only $\sim \theta_c$ away from the 
experimental bound.  For example, Cabibbo haze also allows for 
the possibility of models in which the initial solar and CHOOZ angles are 
similar, which may be an interesting avenue for model building.

We also point out out that the CHOOZ angle is expected to be close to the 
experimental bound within this class of models, 
since Cabibbo effects have a characteristic size.  (Note that this is 
true for neutrino anarchy models in general, though without additional 
effects such as Cabibbo haze such models are somewhat disfavored even with 
the current bound.)  Future experiments, which will either measure 
$\theta_{13}$ or push the bound to 
the ${\cal O}(\lambda^2)$ region, will be crucial in determining whether 
such models remain viable options for flavor model building.

\subsubsection*{One large angle: Cabibbo shifted solar angle.}
In this class of models, $\mathcal{W}$ is assumed to contain only the 
large initial atmospheric angle $\eta_{\oplus}$.  This scenario has 
attractive features from the perspective of 
flavor model building, as it is relatively easy to obtain a structure of 
this type in the context of three family mixing (for example, it 
occurs for mass matrices with a 
degenerate row or column).
In contrast, it is considerably more difficult to generate a pattern of 
two large angles and one small angle without fine-tuning in three family 
models (see e.g. \cite{datta} for discussions of this point). 

However, the price to pay in this scenario is that the 
solar angle must experience a large shift in order to be consistent with 
the data.  Given that this large shift is larger than the characteristic 
size of the perturbations, this scenario requires that the solar angle 
receives several Cabibbo shifts which then sum to a large effect. 
Assuming $\mathcal{V}(\lambda)$ 
of the form given by Eq.~(\ref{vtolamsq}), the shifts are given by \\

\noindent $\bullet$ {\it Right shifts}: $\mathcal{U}_{\rm
MNSP}=\mathcal{R}_1(\eta_\oplus)\mathcal{P}\mathcal{V}(\lambda)$.
\bea
\label{1rcssol}
\theta_\odot&=&\lambda |a_1| +\mathcal{O}(\lambda^2)\\
\label{1rcsatm}
\theta_\oplus&=&\eta_\oplus+\lambda |b_1|\cos(\alpha_{23}+\phi_{b_1}) 
+\mathcal{O}(\lambda^2)\\
\label{1rcschooz}
\theta_{13}&=&\lambda |c_1|+\mathcal{O}(\lambda^2). 
\eea

\noindent $\bullet$ {\it Left shifts}:  $\mathcal{U}_{\rm
MNSP}=\mathcal{V}(\lambda)\mathcal{R}_1(\eta_\oplus)\mathcal{P}$.
\bea
\label{1lcssol}
\theta_\odot&=&\lambda |a_1 \cos\eta_\oplus 
-c_1\sin\eta_\oplus| 
+\mathcal{O}(\lambda^2)\\
\label{1lcsatm}
\theta_\oplus&=&\eta_\oplus+\lambda|b_1|\cos\phi_{b_1} 
+\mathcal{O}(\lambda^2)\\
\label{1lcschooz}
\theta_{13}&=&\lambda|a_1 \sin\eta_{\oplus}+ 
c_1\cos\eta_\oplus |+\mathcal{O}(\lambda^2).
\eea
Eqs.~(\ref{1rcssol}) and (\ref{1lcssol}) show
that it is difficult to obtain a large enough shift in solar angle 
with ${\cal O}(1)$ coefficients. This scenario thus requires either 
large coefficients, which is contrary to the philosophy of treating 
$\lambda$ as an expansion parameter, or more complicated scenarios such as 
sequential perturbations $\mathcal{V}\sim \mathcal{V}_1\mathcal{V}_2$ (in 
which both $\mathcal{V}_1$ and $\mathcal{V}_2$ are at least double shift 
models). The latter approach leads to the following additional 
possibilities:\\ \\ 
\noindent $\bullet$ {\it Right shifts}: $\mathcal{U}_{\rm
MNSP}=\mathcal{R}_1(\eta_\oplus)\mathcal{P}\mathcal{V}_1(\lambda)
\mathcal{V}_2(\lambda)$.
\bea
\label{1drcssol}
\theta_{\odot} & = & \lambda|a_1+a^\prime_1|+\mathcal{O}(\lambda^2)
\\
\label{1drcsatm}
\theta_{\oplus} & = & \eta_{\oplus}+ 
\lambda(|b_1|\cos(\alpha_{23}+\phi_{b_1})+ 
|b'_1|\cos(\alpha_{23}+\phi_{b'_1}))
+\mathcal{O}(\lambda^{2})\\
\label{1drcschooz}
\theta_{13} & = & 
\lambda|c_1+c'_1|+\mathcal{O}(\lambda^{2}).
\eea

\noindent $\bullet$ {\it Left shifts}: $\mathcal{U}_{\rm
MNSP}=\mathcal{V}_1(\lambda) 
\mathcal{V}_2(\lambda)\mathcal{R}_1(\eta_\oplus)\mathcal{P}$.
\bea 
\label{1dlcssol}
\theta_{\odot} & = & \lambda |(a_1+a_1')\cos\eta_{\oplus}-
(c_1+c_1')\sin\eta_{\oplus}| +\mathcal{O}(\lambda^{2})
\\
\label{1dlcsatm}
\theta_{\oplus} & = & 
\eta_{\oplus}+\lambda (|b_1|\cos\phi_{b_1}+ 
|b_1'|\cos\phi_{b'_1})+\mathcal{O}(\lambda^{2})\\
\label{1dlcschooz}
\theta_{13} & = & 
\lambda |(c_1+c_1')\cos\eta_{\oplus}+(a_1+a_1')\sin\eta_{\oplus}|
+\mathcal{O}(\lambda^{2})
\eea

\noindent $\bullet$ {\it Middle shifts}: 
$\mathcal{U}_{\rm 
MNSP}=\mathcal{V}_1(\lambda)\mathcal{R}_1(\eta_\oplus) 
\mathcal{P}\mathcal{V}_2(\lambda)$.
\bea
\label{1dmcssol}
\theta_{\odot} & = & 
\lambda |e^{i\alpha_{12}}a_1'+a_1\cos\eta_{\oplus}-c_1\sin\eta_{\oplus}|
+\mathcal{O}(\lambda^{2})\\
\label{1dmcsatm}
\theta_{\oplus} & = & 
\eta_{\oplus}+\lambda(|b_1|\cos\phi_{b_1}+
|b_1'|\cos(\alpha_{23}+\phi_{b_1'}))
+\mathcal{O}(\lambda^{2})\\
\label{1dmcschooz}
\theta_{13} & = & \lambda 
|e^{i(\alpha_{12}-\alpha_{23})}c_1'+c_1\cos\eta_{\oplus}+
a_1\sin\eta_{\oplus}|+\mathcal{O}(\lambda^{2}).
\eea
These middle shifts are hybrid scenarios (a left shift 
of $\mathcal{V}_1$ and a right shift of $\mathcal{V}_2$), which are
different than the middle shifts discussed previously for 
models with two or more large initial angles and $\mathcal{V}=\mathcal{V}_1$. 

In this class of models the primary issue is to achieve the necessary 
shift in the 
solar angle to boost it into the experimentally measured range.  Hence, 
it is of interest to examine the correlations between this large shift and 
the shifts in the other mixing angles. The prediction for the CHOOZ angle 
is of particular interest, since it must receive a smaller 
Cabibbo shift than the solar angle.   For right shifts, 
Eqs.~(\ref{1drcssol})--(\ref{1drcschooz}) demonstrate 
that the shifts in the mixing angles are uncorrelated at leading order. 
The shifts of the solar and CHOOZ angles are correlated for both left and 
middle shifts (though the atmospheric angle is not), but with a tendency 
toward anticorrelation so these remain viable scenarios. 

From the perspective of the data, this class of models is arguably the 
least plausible Cabibbo haze scenario of the several we have considered.  
In general, shifting the solar angle to its relatively large 
experimental value through effects of Cabibbo size places nontrivial 
constraints on the nature and details of the allowed perturbations, 
although it may still be a worthwhile avenue of exploration for flavor 
model building.

\subsection{CP violation} 
In a three-family mixing scheme, it is well known that if neutrinos are 
Majorana fermions, CP violation in the 
lepton mixing matrix can occur not only from a Dirac phase $\delta_{\rm 
MNSP}$, which is in direct analogy with the quark mixing phase 
$\delta_{\rm CKM}$, but also from two Majorana phases.  The Majorana 
phases do not contribute to the JDGW invariant defined in 
Eq.~(\ref{jarlskogdef}), but their effects can 
be encoded through two additional rephasing invariants chosen from the 
following set \cite{Nieves:1987pp}:
\be
\mathcal{S}_{{\rm CP} \alpha i j}
={\rm 
Im}(\mathcal{U}_{\alpha\,i}\mathcal{U}^*_{\alpha\,j}\xi_j^*\xi_i),  
\ee
in which the $\xi_i$ are given by a generalized Majorana condition
\be
C\bar{\nu}_j^T=\xi^{*2}_j\nu_j.
\ee
Majorana phases only contribute to lepton number violating 
processes such as neutrinoless double beta decay and neutrino-antineutrino 
oscillations (see e.g. \cite{Bilenky:2001rz,deGouvea:2002gf} for 
discussions). These processes are helicity suppressed and difficult to 
observe.  From the standpoint 
of our phenomenological approach, the dependence on the neutrino masses 
$m_i$ (via the helicity suppression factors) adds a 
new facet to the analysis, in that $m_i$ also may be governed by a 
$\lambda$ expansion. Note that as our analysis has been 
independent of the masses so far, we have not specified any details 
about their properties in the $\lambda\rightarrow 0$ limit.  For these 
reasons, we defer such an analysis to a future study and focus instead in 
this paper on Dirac CP violation.   

For Dirac CP violation, the questions to be addressed in our 
phenomenological approach include how the phase $\delta_{\rm MNSP}$ is 
generated in $\mathcal{U}_{\rm MNSP}$ and what are the resulting 
predictions for the lepton sector JDGW invariant  
\be
\mathcal{J}_{\rm CP}\simeq \sin 2\theta_{\oplus}\sin 2\theta_\odot 
\sin 2\theta_{13}\sin\delta_{\rm MNSP}.
\label{delmnspdef}
\ee
The answers hinge upon the nature of CP violation in the 
$\lambda\rightarrow 0$ limit.  As previously mentioned, the starting 
matrix $\mathcal{W}$ may have a number of Dirac phases. 
These phases are not physical in the $\lambda\rightarrow 0$ limit if any 
of the bare mixing angles are zero (see Eq.~(\ref{jarlskogdef})). If 
$\mathcal{W}$ has three large angles, one phase combination 
$\chi$ is physical in the absence of the perturbations, in which case 
$\delta_{\rm MNSP}\sim \chi$ at leading order.  However, even the phases 
which are unobservable in the $\lambda\rightarrow 0$ 
limit may provide a source of CP violation once Cabibbo effects are 
switched on and all three mixing angles are nonvanishing.

For simplicity, and given that our phenomenological approach does not 
address the origin of the physics of the $\lambda\rightarrow 0$ limit, we 
will not elaborate further on scenarios in which $\mathcal{W}$ contains 
bare Dirac phases. We focus instead on cases where these phases are 
absent, but allow for bare Majorana phases and phases in the 
Cabibbo-sized perturbations. In this case, there are two possible ways to 
generate $\delta_{\rm MNSP}$:
\begin{itemize}
\item {\it Generating $\delta_{\rm MNSP}$ from complex 
$\mathcal{V}(\lambda)$.}
In this case, $\mathcal{V}(\lambda)$ is the source of CP-violating phases.   
We assume that these phases can naturally be ${\cal O}(1)$ (as in the 
CKM). Models in this class can 
be categorized in terms of whether CP violation enters at leading or 
higher order in $\lambda$, and whether the effective MNSP phase is 
predicted to be ${\cal O}(1)$ or further suppressed.  

\item {\it  Generating $\delta_{\rm MNSP}$ from Majorana phases.}
Majorana phases can also provide a source for Dirac CP violation once the 
Cabibbo-sized perturbations are switched on in particular scenarios.  For 
left and middle Cabibbo shifts, $\mathcal{P}$ does not contribute to the 
JDGW invariant (by inspection). However, Majorana phases do contribute to 
the JDGW invariant for right Cabibbo shifts, as such shifts encode 
$\mathcal{P}$ through the modification $\mathcal{V} \rightarrow 
\mathcal{V}_{\mathcal{M}}$ as shown in Eq.~(\ref{modrightshifts}).
\end{itemize}

\noindent {\bf Example: Revisiting $\mathcal{U}_{\rm 
MNSP}=\mathcal{U}^\dagger_{\rm CKM}\mathcal{F}$.}\\

\noindent To illustrate these points, we now present a representative 
example of theoretical interest, in which $\mathcal{W}$ has two large 
angles and the hierarchical structure of $\mathcal{V}(\lambda)$ is similar 
to that of the CKM matrix.\footnote{See also \cite{Antusch:2005kw} for 
examples in which the MNSP phase enters directly in the Cabibbo haze
correction to the solar angle.} Specifically, we focus here on the model 
with 
$\mathcal{U}_{\rm MNSP}=\mathcal{U}^\dagger_{\rm CKM}\mathcal{F}$ 
(see Eq.~(\ref{udagf}) and surrounding discussion). This model 
was also discussed in \cite{Datta:2005ci}, but we will do so in 
slightly more general terms here.  First recall that in this model, 
which is a 
left shift scenario with $\mathcal{V}=\mathcal{U}^\dagger_{\rm CKM}$ 
and 
$\mathcal{F}=\mathcal{W}=\mathcal{R}_1(\eta_\oplus)\mathcal{R}_3(\eta_\odot)\mathcal{P}$, 
the shifts in the angles are given by
\bea 
\theta_\odot&=&\eta_\odot-\lambda\cos\eta_\oplus+\mathcal{O}(\lambda^3)\\ 
\theta_\oplus&=&\eta_\oplus-\lambda^2(A+\frac{1}{4}\sin 2\eta_\oplus) + 
\mathcal{O}(\lambda^3)\\
\theta_{13}&=&-\lambda\sin\eta_\oplus+\mathcal{O}(\lambda^3). 
\eea
As this model is a left shift scenario, the dominant shift in 
$\theta_{13}$ is $\sim \lambda$ due to the ${\cal O}(\lambda)$ mixing 
between the first and second families in the quark sector (in addition, 
the Majorana phases do not enter the Cabibbo shifts).  
  
Using the Wolfenstein form of $\mathcal{U}_{\rm CKM}$ as given in 
Eq.~(\ref{wolf}), the JDGW invariant for $\mathcal{U}_{\rm MNSP}$ is 
(see also \cite{Datta:2005ci}):
\be
\mathcal{J}_{\rm CP}=\frac{1}{4}A\lambda^3\eta \cos\eta_\oplus\sin 
2\eta_\oplus \sin 2\eta_\odot.
\label{udagfj}
\ee
Given that this model has one ${\cal O}(\lambda)$ and two ${\cal O}(1)$ 
mixing angles, Eq.~(\ref{udagfj}) indicates that 
$\delta_{\rm MNSP}\sim \mathcal{O}(\lambda^2)$, in contrast to the 
${\cal O}(1)$ CKM phase $\delta_{\rm CKM}$.  This suppression occurs 
because the phases in $\mathcal{V}$ are only manifest in subdominant 
contributions to the mixing angles.  As discussed in \cite{Datta:2005ci}, 
models with this feature demonstrate that while the size of $\theta_{13}$ 
is clearly correlated with the prospects for the observability of lepton 
sector CP violation, it does not tell the whole story because the 
CP-violating phase itself may be additionally suppressed.

Eq.~(\ref{udagfj}) was obtained using a very specific form of 
$\mathcal{U}_{\rm CKM}$, in which the freedom to rotate away 
phases which are unphysical in the Standard Model has already been taken 
into account and the remaining phase degree of freedom was chosen to appear at 
a particular location in the CKM matrix.  For the quarks, this  
does not lead to any loss of generality as the JDGW invariant is 
independent of this choice by construction.  However, it is 
too preliminary to restrict the CKM to this form when computing the  
JDGW invariant for the leptons within this class of models, and 
indeed the location of the CP-violating phase(s) affects the 
results.  To see this 
more clearly, let us write the CKM in the form given in 
Eqs.~(\ref{aexpr})--(\ref{vtolamsq}), leaving the coefficients $a_1$, 
$b_2$, and $c_3$ general rather than specifying the Wolfenstein values. 
While the JDGW invariant for $\mathcal{U}^\dagger_{\rm CKM}$ is
\be
\mathcal{J}^{\rm (CKM^\dagger)}_{\rm CP}=-\lambda^6 {\rm Im}(a_1b_2c^*_3),
\ee 
which shows the convention independence explicitly, the JDGW 
invariant for $\mathcal{U}_{\rm MNSP}=\mathcal{U}^\dagger_{\rm 
CKM}\mathcal{F}$ is 
\bea
\mathcal{J}_{\rm CP}&=&\lambda{\rm Im}(a_1)
\sin\eta_\oplus\sin 2\eta_\oplus\sin 2\eta_\odot 
(1-\frac{\lambda^2}{4}|a_1|^2)+\frac{\lambda^3}{4}{\rm 
Im}(c_3)\cos\eta_\oplus \sin 2\eta_\oplus\sin 2\eta_\odot
\nonumber \\
&+&\frac{\lambda^3}{8}{\rm Im}(a_1b_2)\sin\eta_\oplus(1-3\cos 
2\eta_\oplus)\sin 2\eta_\odot-\frac{\lambda^3}{4}{\rm Im}(a_1b_2^*) 
\cos\eta_\oplus\sin 2\eta_\oplus \sin 2\eta_\odot. 
\label{udagfgenj}
\eea
The lepton JDGW invariant is ${\cal O}(\lambda^3)$ if ${\rm Im}(a_1)=0$ 
and either ${\rm Im}(b_2)\neq 0$ or ${\rm Im}(c_3)\neq 0$ (note that 
Eq.~(\ref{udagfgenj}) 
reduces to Eq.~(\ref{udagfj}) for the Wolfenstein parameters ${\rm 
Im}(a_1)=0$, ${\rm Im}(b_2)=0$, ${\rm Im}(c_3)=-A\eta$). However, if ${\rm 
Im}(a_1)\neq 0$, the JDGW invariant is ${\cal O}(\lambda)$ and the 
effective $\delta_{\rm MNSP}$ is ${\cal O}(1)$, because in this case a 
CP-violating phase enters in the dominant shift of the CHOOZ angle.  
Hence, whether the JDGW invariant is ${\cal O}(\lambda)$ or ${\cal 
O}(\lambda^3)$ in this model will depend on the details of the associated 
model for $\mathcal{U}_{\rm CKM}$. 

The lesson to be learned from this exercise is a point which was 
also stressed in \cite{Datta:2005ci}: the location of the phases in 
$\mathcal{V}$, and whether they enter in the dominant or subdominant 
shifts in the mixing angles, plays an important role in determining the 
size of Dirac CP violation.  With this in mind, we now turn to the a 
discussion of the JDGW invariants for the different Cabibbo haze 
scenarios.  For economy of presentation, we present the leading order in 
$\lambda$ results for each scenario of interest and do not display higher 
order contributions.  Within our range of assumptions, this corresponds to 
scenarios in which $\delta_{\rm MNSP}$ is ${\cal O}(1)$ in most cases.  
The exception is the class of models with three large initial mixing 
angles, in which case the leading order effects correspond to $\delta_{\rm 
MNSP}\sim {\cal O}(\lambda)$ since we have explicitly chosen not 
to consider situations in which there are bare Dirac phases.

\subsubsection*{Two large angles.}
Once again, we begin with the Cabibbo haze scenario in 
which $\mathcal{W}$ has two large angles, which was also studied in the 
absence of bare Majorana phases in \cite{Datta:2005ci} (to which we refer 
the reader for details).\\

\noindent $\bullet$ {\it Right shifts}: $\mathcal{U}_{\rm
MNSP}=\mathcal{R}_1(\eta_\oplus)\mathcal{R}_3(\eta_\odot) 
\mathcal{P}\mathcal{V}(\lambda)$.
\be
\label{jdgw2r}
\mathcal{J}_{\rm CP}=-\frac{1}{4}\lambda \sin2\eta_\oplus\sin 
2\eta_\odot(\sin\eta_\odot|b_1|\sin(\alpha_{23}+\phi_{b_1})
+\cos\eta_\odot|c_1|\sin(\alpha_{12}-\alpha_{23}+\phi_{c_1}))+
\mathcal{O}(\lambda^2).
\ee
\noindent $\bullet$ {\it Left shifts}:  $\mathcal{U}_{\rm
MNSP}=\mathcal{V}(\lambda)\mathcal{R}_1(\eta_\oplus)\mathcal{R}_3(\eta_\odot)
\mathcal{P}$.

\be
\label{jdgw2l}
\mathcal{J}_{\rm CP}=-\frac{1}{4}\lambda \sin2\eta_\oplus\sin 
2\eta_\odot(\sin\eta_\oplus|a_1|\sin\phi_{a_1}+ 
\cos\eta_\oplus|c_1|\sin\phi_{c_1})
+\mathcal{O}(\lambda^2).
\ee
\noindent $\bullet$  {\it Middle shifts}: $\mathcal{U}_{\rm
MNSP}=\mathcal{R}_1(\eta_\oplus)\mathcal{V}(\lambda)
\mathcal{R}_3(\eta_\odot)\mathcal{P}$.

\be
\label{jdgw2m}
\mathcal{J}_{\rm CP}=-\frac{1}{4}\lambda 
\sin2\eta_\oplus
\sin 2\eta_\odot|c_1|\sin\phi_{c_1}
+\mathcal{O}(\lambda^2).
\ee
The leading order contributions to the JDGW invariant are ${\cal 
O}(\lambda)$, which is to be expected since the CHOOZ angle is by 
construction a Cabibbo effect in this class of models.  Note that as 
expected, the right shifts depend on the Majorana phases, allowing  
for an ${\cal O}(\lambda)$ CP violation even if the leading order 
perturbations do not explicitly involve CP-violating phases.
In all cases, the effective $\delta_{\rm MNSP}$ is ${\cal O}(1)$ unless 
these leading order contributions vanish for any given scenario, in which 
case $\delta_{\rm MNSP}$ is further suppressed.  We refer the reader to 
\cite{Datta:2005ci} for examples of each of these scenarios and additional 
discussion.  

\subsubsection*{Three large angles.}
We now turn to the case of three large angles, keeping in mind that 
since we do not consider the possibility of bare Dirac CP-violating 
phases, the leading order contributions to the JDGW invariant 
are of ${\cal O}(\lambda)$.  They are given 
for each scenario as follows:\\

\noindent $\bullet$ {\it Right 
shifts}: 
$\mathcal{U}_{\rm MNSP}=\mathcal{R}_1(\eta_\oplus)\mathcal{R}_2(\eta_{13})
\mathcal{R}_3(\eta_\odot)\mathcal{P}\mathcal{V}(\lambda)$.
\bea
\label{jdgw3r}
\mathcal{J}_{\rm CP} & = & 
\frac{1}{2}\lambda[|a_1|\sin2\eta_{\oplus}\sin\eta_{13} 
\cos^{2}\eta_{13}\sin(\alpha_{12}+\phi_{a_1})+\\
 & & 
|b_1|\sin(\alpha_{23}+\phi_{b_1})\cos\eta_{13}\cos\eta_{\odot} 
(\sin2\eta_{\oplus}(\cos^{2}\eta_{\odot}\sin^{2}\eta_{13}- 
\sin\eta_{\odot}^{2})
+\sin\eta_{13}\sin2\eta_{\odot}\cos2\eta_{\oplus})+\nonumber\\
 & & 
|c_1|\sin(\alpha_{12}-\alpha_{23}+ \phi_{c_1})\cos\eta_{13} 
\sin\eta_{\odot}(\sin2\eta_{\oplus}(\sin^{2}\eta_{\odot}
\sin^{2}\eta_{13}-\cos\eta_{\odot}^{2})-
\sin\eta_{13}\sin2\eta_{\odot}\cos2\eta_{\oplus})]\nonumber 
\\
&&+\mathcal{O}(\lambda^2).\nonumber
\eea

\noindent $\bullet$  {\it Left shifts}: 
 $\mathcal{U}_{\rm
MNSP}=\mathcal{V}(\lambda)
\mathcal{R}_1(\eta_\oplus)\mathcal{R}_2(\eta_{13})
\mathcal{R}_3(\eta_\odot)\mathcal{P}$.
\bea
\label{jdgw3l}
\mathcal{J}_{\rm CP} & = & 
-\frac{1}{2}\lambda \cos\eta_{13}
[|a_1|\sin\phi_{a_1}\cos\eta_{13}\cos\eta_{\oplus}\sin2\eta_{\odot}- 
\frac{1}{2}|b_1|\sin\phi_{b_1}\sin2\eta_{13}\sin2\eta_\odot \nonumber\\& 
& 
-(|a_1|\sin\phi_{a_1}\cos\eta_\oplus-|c_1|\sin\phi_{c_1}\sin\eta_\oplus) 
(\cos2\eta_\odot\sin2\eta_\oplus\sin\eta_{13}+
\sin2\eta_\odot(\cos\eta_\oplus^2-\sin\eta_{13}^2\sin\eta_\oplus^2))]
\nonumber\\&&
+\mathcal{O}(\lambda^2).
\eea

\noindent $\bullet$ {\it Middle left shifts}:  $\mathcal{U}_{\rm
MNSP}=\mathcal{R}_1(\eta_\oplus)\mathcal{V}(\lambda)\mathcal{R}_2(\eta_{13})
\mathcal{R}_3(\eta_\odot)\mathcal{P}$.

\bea
\label{jdgw3ml}
\mathcal{J}_{\rm CP} & = & 
\frac{1}{2}\lambda\cos\eta_{13}[
|a_1|\sin\phi_{a_1}\sin\eta_{13}(\sin2\eta_\odot 
\cos2\eta_\oplus\sin\eta_{13}+\cos2\eta_\odot\sin2\eta_\oplus)\nonumber \\
&&+\frac{1}{2}(|b_1|\sin\phi_{b_1} 
\sin2\eta_\odot\sin2\eta_{13}\cos2\eta_\oplus - 
|c_1|\sin\phi_{c_1}\cos\eta_{13}^2\sin2\eta_\odot\cos2\eta_\oplus)]
+\mathcal{O}(\lambda^2).
\eea

\noindent $\bullet$  {\it Middle right shifts}: 
$\mathcal{U}_{\rm 
MNSP}=\mathcal{R}_1(\eta_\oplus)\mathcal{R}_2(\eta_{13}) 
\mathcal{V}(\lambda)\mathcal{R}_3(\eta_\odot)\mathcal{P}$.
\bea
\label{jdgw3mr}
\mathcal{J}_{\rm CP} & = & 
\frac{1}{2}\lambda\cos\eta_{13}[\frac{1}{2} 
|a_1|\sin\phi_{a_1}\sin2\eta_{13}\sin2\eta_{\oplus} 
\cos2\eta_{\odot}+
|b_1|\sin\phi_{b_1}\sin\eta_{13}(\cos2\eta_\odot\sin2\eta_\oplus\sin\eta_{13}\nonumber 
\\&& 
+\sin2\eta_\odot(\cos\eta_\oplus^2-\sin\eta_{13}^2\sin\eta_\oplus^2)) 
- \cos\eta_{13}^2\sin2\eta_\odot\sin\eta_\oplus 
(|b_1|\sin\phi_{b_1}\sin\eta_{13}\sin\eta_\oplus+ 
|c_1|\sin\phi_{c_1}\cos\eta_\oplus)]\nonumber\\
&&+\mathcal{O}(\lambda^2).
\eea
Note that Eqs.~(\ref{jdgw3r})--(\ref{jdgw3mr}) reduce to their appropriate 
counterparts of Eqs.~(\ref{jdgw2r})--(\ref{jdgw2m}) in the limit that 
$\eta_{13}\rightarrow 0$.

\subsubsection*{One large angle.}
Finally, we present the leading order JDGW invariants for the class of 
models in which the bare solar and CHOOZ angles are zero.  These 
contributions are of ${\cal O}(\lambda^2)$, which is expected since two of 
the three MNSP mixing angles are due to Cabibbo effects, and therefore 
correspond to an effective $\delta_{\rm MNSP}$ of ${\cal O}(1)$.  For 
perturbations given by Eq.~(\ref{vtolamsq}), the JDGW invariants are:\\

\noindent $\bullet$ {\it Right shifts}:  $\mathcal{U}_{\rm
MNSP}=\mathcal{R}_1(\eta_\oplus)\mathcal{P}\mathcal{V}(\lambda)$.
\be
\label{jdgw1r}
\mathcal{J}_{\rm CP}=-\frac{1}{2}\lambda^2 \sin2\eta_\oplus
|a_1c_1|\sin(\alpha_{23}-\phi_{a_1}+\phi_{c_1})
+\mathcal{O}(\lambda^3).
\ee
\noindent $\bullet$  {\it Left shifts}: $\mathcal{U}_{\rm
MNSP}=\mathcal{V}(\lambda) \mathcal{R}_1(\eta_\oplus)\mathcal{P}$.

\be
\label{jdgw1l}
\mathcal{J}_{\rm CP}=\frac{1}{2}\lambda^2 
|a_1c_1|\sin(\phi_{a_1}-\phi_{c_1})
+\mathcal{O}(\lambda^3).
\ee
We also present the JDGW invariants for perturbations of the form 
$\mathcal{V}=\mathcal{V}_1\mathcal{V}_2$:\\

\noindent $\bullet$ {\it Right shifts}: $\mathcal{U}_{\rm
MNSP}=\mathcal{R}_1(\eta_\oplus)\mathcal{P}\mathcal{V}_1(\lambda) 
\mathcal{V}_2(\lambda)$.

\bea
\label{jdgw1dr}
\mathcal{J}_{\rm CP}&=&-\frac{1}{2}\lambda^2 \sin2\eta_\oplus
[ |a_1c_1|\sin(\alpha_{23}-\phi_{a_1}+\phi_{c_1})+
|a_1'c_1|\sin(\alpha_{23}-\phi_{a_1'}+\phi_{c_1})\nonumber \\&&+ 
|a_1c_1'|\sin(\alpha_{23}-\phi_{a_1}+\phi_{c_1'})+
|a_1'c_1'|\sin(\alpha_{23}-\phi_{a_1'}+\phi_{c_1'}) ]
+\mathcal{O}(\lambda^3).
\eea
\noindent $\bullet$ {\it Left shifts}: $\mathcal{U}_{\rm
MNSP}=\mathcal{R}_1(\eta_\oplus)\mathcal{P}\mathcal{V}_1(\lambda) 
\mathcal{V}_2(\lambda)$.
\bea
\label{jdgw1dl}
\mathcal{J}_{\rm CP}&=&\frac{1}{2}\lambda^2 
[ |a_1c_1|\sin(\phi_{a_1}-\phi_{c_1})- 
|a_1'c_1|\sin(\phi_{c_1}-\phi_{a_1'})+\nonumber \\ &&
|a_1c_1'|\sin(\phi_{a_1}-\phi_{c_1'})+|a_1'c_1'|\sin(\phi_{a_1'}-\phi_{c_1'}) 
]
+\mathcal{O}(\lambda^3).
\eea
\noindent $\bullet$  {\it Middle shifts}: $\mathcal{U}_{\rm
MNSP}=\mathcal{V}_1(\lambda) \mathcal{R}_1(\eta_\oplus)\mathcal{P}
\mathcal{V}_2(\lambda)$. 
\bea
\label{jdgw1dm}
\mathcal{J}_{\rm CP}&=&\frac{1}{2}\lambda^2 \sin 2\eta_\oplus 
[ \sin\eta_\oplus (|a_1a_1'|\sin(\alpha_{12}-\phi_{a_1}+\phi_{a_1'}) 
+|c_1c_1'|\sin(\alpha_{12}-\alpha_{23}-\phi_{c_1}+\phi_{c_1'}))+ \nonumber 
\\&& 
\cos\eta_\oplus(|c_1a_1'|\sin(\alpha_{12}-\phi_{c_1}+\phi_{a_1'}) - 
|a_1c_1'|\sin(\alpha_{12}-\alpha_{23}-\phi_{a_1}+\phi_{c_1'}))+ \nonumber 
\\&&
|a_1c_1|\sin(\phi_{a_1}-\phi_{c_1})-|a_1'c_1'|\sin(\phi_{a_1'}-\phi_{c_1'})] 
+\mathcal{O}(\lambda^3).
\eea
Note that for such sequential perturbations, the JDGW invariants for the 
middle as well as right shifts depends on the bare Majorana phases.  This 
is easily understood because these middle shift scenarios are a hybrid of 
right shifts (which involve the Majorana phases) and left shifts 
(which don't).

We conclude by stressing a point mentioned in 
\cite{Datta:2005ci} which is worth repeating here.  Our catalog focuses on 
the leading order contributions to the JDGW invariants, which 
usually lead to an ${\cal O}(1)$ $\delta_{\rm MNSP}$ (similar to 
$\delta_{\rm CKM}$).  However, since leading 
order contributions can be absent, it is not automatic that the MNSP 
phase is a large angle.  Many models lead to a suppressed 
$\delta_{\rm MNSP}$; for example, this occurs for left and middle 
shifts (and right shifts if bare Majorana phases vanish) if the 
dominant shifts are manifestly real.  
In such cases, a nonvanishing CHOOZ angle is a necessary but not 
sufficient condition for the prospects for observing leptonic CP 
violation.
 
\section{Conclusions} 
The recent experimental progress in the lepton sector has 
revealed new avenues for exploration in the search to 
formulate a compelling theory of the masses and mixings of the SM 
fermions.  As a step toward this elusive goal, we have advocated a 
phenomenological approach which proposes that both lepton and quark 
mixings can be understood as expansions in the Cabibbo angle $\lambda$.  
This hypothesis can be argued in the context of grand unification, 
and more generally in theoretical frameworks in which the flavor 
structures of the quarks and leptons are controlled by the same order 
parameter.  Within this approach, the lepton mixings are enveloped in a 
haze of Cabibbo-sized effects, as (unlike the quark mixings) they are 
unknown in the $\lambda\rightarrow 0$ limit.

To aid in viewing lepton mixing through the Cabibbo haze, this paper 
provides a systematic classification of possible parametrizations which 
includes a catalog of the ${\cal O}(\lambda)$ effects on the mixing angles 
and the CP-violating phases. Although present experimental 
constraints are not sufficient to single out a particular parametrization, 
this phenomenological approach has practical applications both in 
categorizing top-down flavor models and in suggesting a roadmap for future 
measurements of the MNSP matrix. Should the limit of zero Cabibbo 
angle prove to be meaningful for theory, with improved data we may be able 
to discern the underlying flavor theory through the 
Cabibbo haze.

\section*{Acknowledgments} We thank D. Chung, L. Duffy, K. Matchev, P. 
Ramond, and C. Thorn for helpful discussions. We also thank S. Petcov 
for the suggestion to include Majorana phases, as well as many other 
helpful comments.  This work is supported by the U. S. Department of 
Energy under grant DE-FG02-97ER41209, and  
and a L'Or\'{e}al for Women in Science Postdoctoral Fellowship.

\end{document}